\documentclass[conference]{IEEEtran}
\pdfoutput=1
\IEEEoverridecommandlockouts
\usepackage{hyperref} 
\usepackage{setspace}
\usepackage{cite}
\usepackage{amsmath,amssymb,amsfonts}
\usepackage{graphicx}
\usepackage{textcomp}
\usepackage{algorithm}
\usepackage[noend]{algpseudocode}

\def\BibTeX{{\rm B\kern-.05em{\sc i\kern-.025em b}\kern-.08em
    T\kern-.1667em\lower.7ex\hbox{E}\kern-.125emX}}

\makeatletter
\def\@IEEEpubidpullup{8\baselineskip}
\makeatother

\begin{document}

\IEEEoverridecommandlockouts
\IEEEpubid{
\parbox{\columnwidth}{\vspace{-4\baselineskip}Permission to make digital or hard copies of all or part of this work for personal or classroom use is granted without fee provided that copies are not made or distributed for profit or commercial advantage and that copies bear this notice and the full citation on the first page. Copyrights for components of this work owned by others than ACM must be honored. Abstracting with credit is permitted. To copy other
wise, or republish, to post on servers or to redistribute to lists, requires prior specific permission and/or a fee. Request permissions from \href{mailto:permissions@acm.org}{permissions@acm.org}.\hfill\vspace{-0.8\baselineskip}\\
\begin{spacing}{1.2}
\small\textit{UbiComp/ISWC '18} Adjunct, October 8–12, 2018, Singapore. \\
\copyright\space2018 Association for Computing Machinery. \\
\$15.00 \\
\url{http://dx.doi.org/xx.xxxx/xxxxxxx.xxxxxxx} 
\end{spacing}
\hfill}
\hspace{0.9\columnsep}\makebox[\columnwidth]{\hfill}}
\IEEEpubidadjcol

\title{Community Structures, Interactions and Dynamics in London's Bicycle Sharing Network}

\author{\IEEEauthorblockN{
Fernando Munoz-Mendez\IEEEauthorrefmark{1}\IEEEauthorrefmark{4}, 
Konstantin Klemmer\IEEEauthorrefmark{2}\IEEEauthorrefmark{4}, 
Ke Han\IEEEauthorrefmark{3} and 
Stephen Jarvis\IEEEauthorrefmark{2}}
\IEEEauthorblockA{\IEEEauthorrefmark{1}AECOM, Bedford, UK\\
Email: fernando.munozmendez@aecom.com}
\IEEEauthorblockA{\IEEEauthorrefmark{2}University of Warwick, Coventry, UK and \\Alan Turing Institute, London, UK\\
Email: \{k.klemmer, s.a.jarvis\}@warwick.ac.uk}
\IEEEauthorblockA{\IEEEauthorrefmark{3}Imperial College London, London, UK\\
Email: k.han@imperial.ac.uk}
\IEEEauthorblockA{\IEEEauthorrefmark{4}\textit{These authors contributed equally to this work.}}
}

\maketitle

\begin{abstract}
Bikesharing schemes are transportation systems that not only provide an efficient mode of transportation in congested urban areas, but also improve last-mile connectivity with public transportation and local accessibility. Bikesharing schemes around the globe generate detailed trip data sets with spatial and temporal dimensions, which, with proper mining and analysis, reveal valuable information on urban mobility patterns. In this paper, we study the London bicycle sharing dataset to explore community structures. Using a novel clustering technique, we derive distinctive behavioural patterns and assess community interactions and spatio-temporal dynamics. The analyses reveal self-contained, interconnected and hybrid clusters that mimic London's physical structure. Exploring changes over time, we find geographically isolated and specialized communities to be relatively consistent, while the remaining system exhibits volatility, especially during and around peak commuting times. By increasing our understanding of the collective behaviour of the bikesharing users, this analysis supports policy appraisal, operational decision-making and motivates improvements in infrastructure design and management.
\end{abstract}


\section{Introduction}
Bicycle Sharing Schemes (BSS) have become increasingly vital elements of urban mobility due to their complementary effect to conventional modes and last-mile connectivity to transit systems \cite{Liu2012}. By now, there are more than 600 BSS globally with the largest systems in China, and successful deployments in Paris, London and Washington D.C. The health benefits of bicycle use in cities even outweighs accident risk \cite{Rojas-Rueda2012}. Furthermore, BSS offer sustainable solutions to urban transportation by contributing to resolving the thriving problems of congestion and pollution. In order to increase the expansion of BSS and attract new customers, it is vital to understand relevant spatial travel patterns, and adjust design and management strategies (e.g. pricing, marketing, expansions) to encourage adoption. For example, if bikesharing is utilized for last mile travel, then a transfer fare could increase its usage and simultaneously promote public transit adoption \cite{Guo2011}. Beyond that, a better understanding of trip patterns will allow for advanced bicycle relocation strategies and more reliable service provision which, in turn, will make the system more attractive to users. The key challenge for any shared mobility system lies in the respective network complexity, noise and the resulting operational implications. To overcome this, we propose a comprehensive and pervasive station-level characterization of the London network, based on spatio-temporal utilization features. Our framework extracts largely self-contained clusters which not only provide insight into mobility patterns, but also help with identifying bottlenecks and inefficiencies, and hence help decision makers to better understand supply and demand imbalances, plan operations and manage infrastructure. Comparing the explicitly non-spatial network model to the known geospatial structure of the system enables us to assess whether communities are a result of space (geography) or place (local features). We develop this approach deploying a dynamic community analysis from BSS rental data collected in London. We introduce a novel approach to community detection in BSS networks and assess interactions between communities as well as the convergence of communities during different hours of the day.

The remainder of this paper is organized as follows. Section 2 reviews relevant studies of BSS and community analysis. A detailed description of the London BSS dataset is provided in Section 3. Section 4 describes the main methodologies and procedures involved in the data-driven analyses. Section 5 contains the discussion and concluding remarks.

\section{Related Work}

With the proliferation of smart data related to BSS there has been a significant amount of research dedicated to either improving our understanding of the BSS to support evidence based policy making, or to perform logistic optimization methods for bicycle relocation. Based on a Barcelona BSS data, \cite{Froehlich2009} applied spatio-temporal analyses, clustering techniques and tested the performance of various machine learning algorithms. \cite{Kaltenbrunner2010} estimates station-level time-series using autoregressive predictive models on the same dataset. Other efforts include a characterization of the network based on the usage profiles of the stations. Vienna’s BSS was analyzed to obtain distinct clusters using partitioning algorithms on usage time-series data in addition to a predictive method to forecast ridership volume \cite{Vogel2011}. BSS stations in Paris were analyzed in respect to usage counts, using a novel Expectation Maximization (EM) model and relating the identified clusters according to their spatial relationships \cite{Etienne2014}. London’s BSS was examined to detect differences in the patterns of usage before and after the opening of the scheme to unregistered (casual) users \cite{Lathia2012} and to identify commuters \cite{Martin-Moral2017}. A big differentiating point between these studies is the type of data used: While some research only has access to availability data at station-level, others are able to use arrival and departure data, detailing every trip. This second set provides more detail as it captures information about periods of inactivity and activity, unlike availability data, which loses information when the net change in bikes at a station is small. 

Another focus of analysis in BSS is the detection of communities, that is, the detection of groups of individual users or stations that exhibit a stronger interdependence between one another, as opposed to other members of the system. This allows for a spatial aggregation of the network and the extraction of patterns. \cite{ZaltzAustwick2013} analyze community structures in five urban BSS \textit{(London, UK; Boston, MA; Denver, CO; Minneapolis, MN; and Washington, DC)}. However, due to limited data availability, the researchers had to generate their own origin / destination (OD) matrices. Furthermore, their approach employs hierarchical community detection, which comes with some shortcomings, especially when resolving the boundaries of different communities or relating nodes that do not share any connections \cite{Newman2004}. \cite{BORGNAT2011a} perform community detection aggregation on the Lyon BSS dataset using the Louvain algorithm. Overall, we observe that existing literature on community structure detection in BSS lacks the use of non modularity based methods and more granular, empirical trip data. From a technical viewpoint, community detection using modularity maximization is known to be vulnerable to resolution restrictions, is limited to undirected information and assumes a process of endogenous network formation \cite{Fortunato2006}. Furthermore, the interaction between extracted communities and the evolution of communities over time remain largely unexplored fields. With our paper, we seek to address these gaps in research.

\section{Data}

\begin{figure}[htbp]
\includegraphics[,scale=0.6]{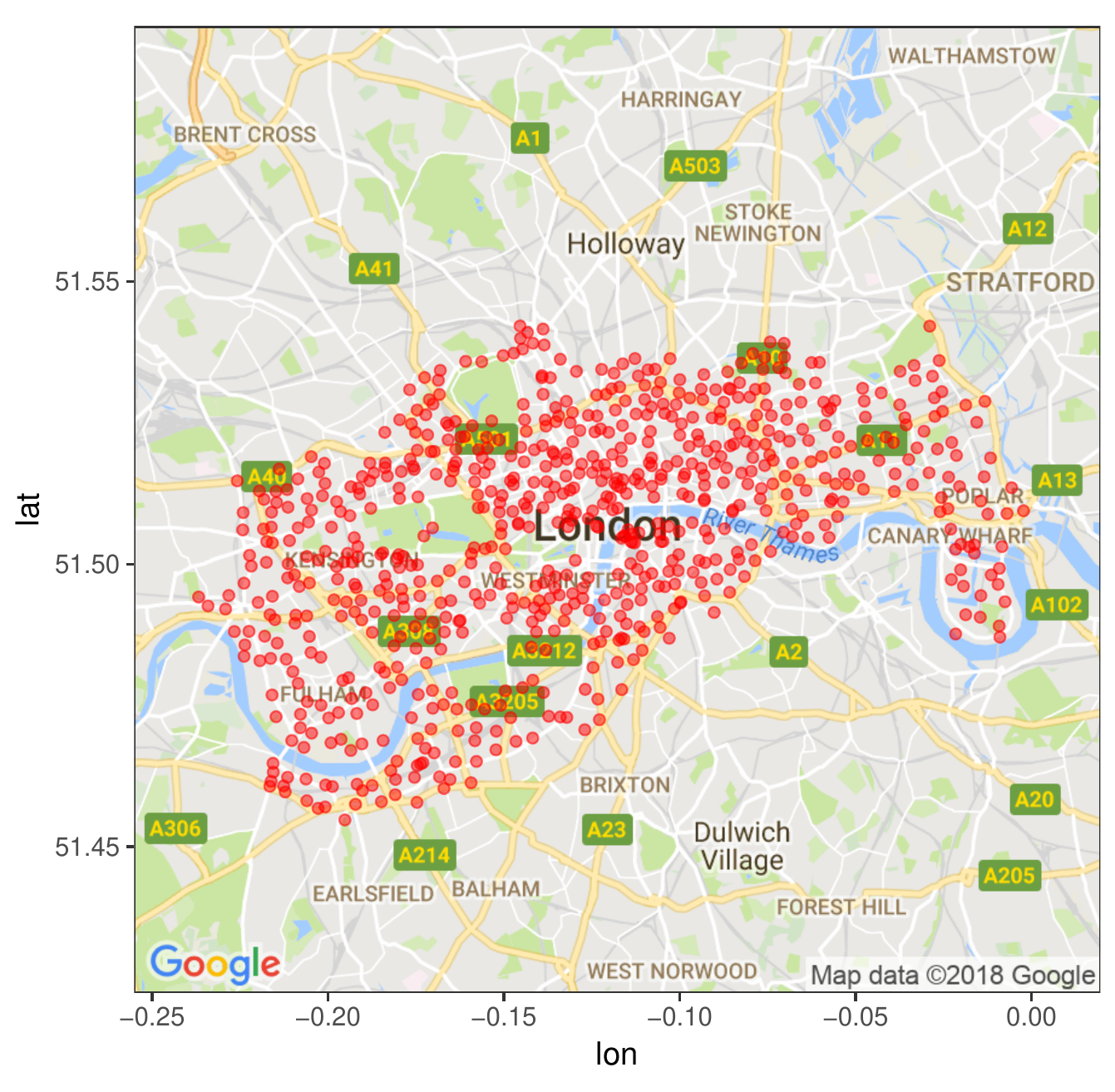}
\caption{Location of bikesharing stations in London}
\label{fig:fig1}
\end{figure} 

The data for our analysis comes from the Transport for London \textit{(TfL)} Open Data API \footnote{Accessible via: https://api.tfl.gov.uk/}and contains information about the unique IDs for each bicycle, the names and IDs of the origin and destination stations, a unique transaction ID (per trip, not per user) and the start and end times of each rental. The dataset covers every recorded shared bicycle trip since 2012 and hence comprises millions of entries. Operating with such large amounts of data can be computationally expensive. Our further analysis will hence utilize an applicable subsample.

First, however, we address some minor problems regarding the collection of trip information, that translates into missing retrieval for destination station IDs or trips without information about bicycle IDs. These issues are not temporally consistent across the dataset, with some months exhibiting higher error rates than others. To adapt the analysis accordingly, we select a small, particularly accurate interval where a crude cleaning of the data, i.e. a removal of incomplete entries, does not result in significant non-response bias. We clean our subset by removing the following entries:
\begin{itemize}
\item Trips that start or end at a repair station.
\item Trips that do not report correct destinations and show a negative duration.
\item Trips that do not report the bicycle ID.
\end{itemize}

\begin{figure*}[htbp]
\centering
\includegraphics[,scale=0.6]{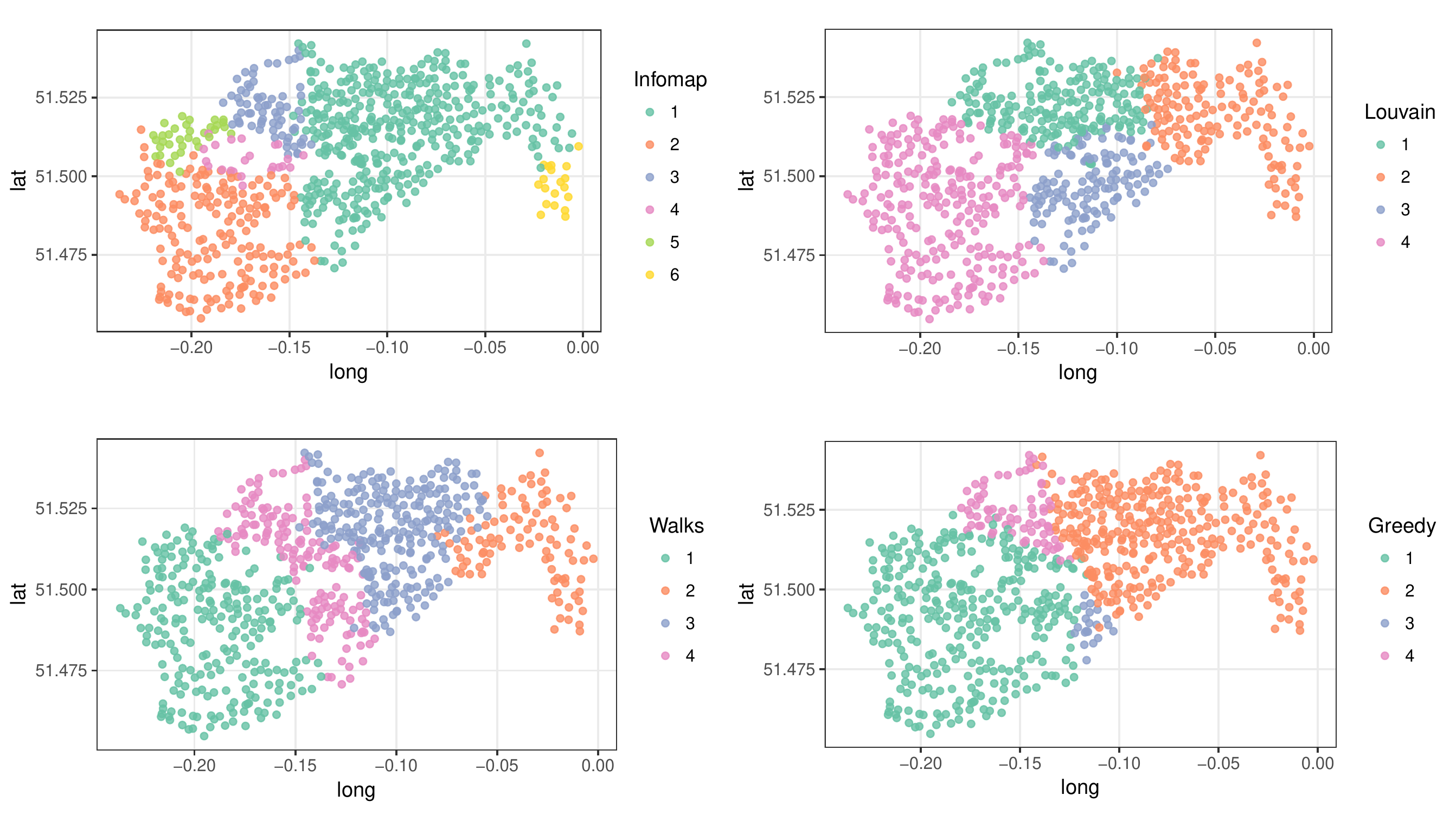}
\caption{Comparison of community detection algorithms: BSS stations are colored according to their respective community assignment across the four techniques}
\label{fig:fig2}
\end{figure*}

Our final dataset is comprised of 1,469,945 unique shared bicycle trips in June and July 2014, distributed over 750 stations in London (see Figure 1). Weekends are disregarded, as their varying trip patterns and added noise could harm our analysis. Aggregated on a station level, our data can be used to compile an Origin-Destination (OD) matrix. We can then formulate a graph $G$ with each station describing a network node $N_{\alpha}$, linked to every other station in the network by a set of directed edges $E_{\alpha}$ weighted by a flow $w_{\alpha}$ equal to the number of trips observed, given in the OD matrix.   

\section{Methodology}

\subsection{Community detection}

Community detection techniques aim at reducing the complexity of a network to a degree that enables comprehensive insight into the underlying network structure. BSS are naturally suited for such applications. Simplifying BSS network descriptions and detecting clusters of stations that exchange many trips also have immediate operational implications by providing a valuable decision-support tool for network management and expansion. Having said that, reliable results depend on an appropriate choice of the community detection algorithm. The most popular methods rely on modularity maximization; however, they do not seem to be applicable in our case, as outlined in section \textit{II}. Most importantly, these methods assume an underlying process of network formation which, in the case of station-based BSS, is not present. This problem also motivated the development of the \textit{Infomap} algorithm, originally proposed by Rosvall \& Bergstrom \cite{Rosvall2008}. This method acknowledges that the system structure drives the flow in the system, leading to system-wide interdependencies. By partitioning the network, the length of the description of the movements can be longer or shorter (bigger and smaller cost of information). By choosing the partition that minimizes the description length, we find the division that provides the best representation of the community structures.

Deploying the \textit{Infomap} approach, we seek to partition the network nodes $N_{1,2,...,n}$ into $M_{1,2,...,m}$ modules by minimizing the information cost of describing the movements of a random walker or, if available, the empirical flow (here the trips) through the network. This is implemented by the \textit{map equation}:

\begin{equation}
L(M) = q_{\curvearrowright}H(\mathcal{Q}) + \sum^{m}_{i=1}p_{\circlearrowright}^{i}H(\mathcal{P}^{i}),
\end{equation}

where $q_{\curvearrowright}$ gives the probability that the random walker leaves the current module, $p_{\circlearrowright}^{i}$ gives the proportion the walker spends in the respective module $\mathcal{P}^{i}$, $H(\mathcal{Q})$ gives the index codebook entropy and $H(\mathcal{P}^{i})$ gives the module codebook entropy\footnote{Referring to Shannon's source coding theorem \cite{Shannon1948}}. Practically, this is carried out by assigning the modules $M_{i}$ to a neighboring module $M_{\beta}$, as long as this reduces $L(M)$. The \textit{Infomap} algorithm can then be applied as follows.

\begin{figure*}[htbp]
\centering
\includegraphics[,scale=0.4]{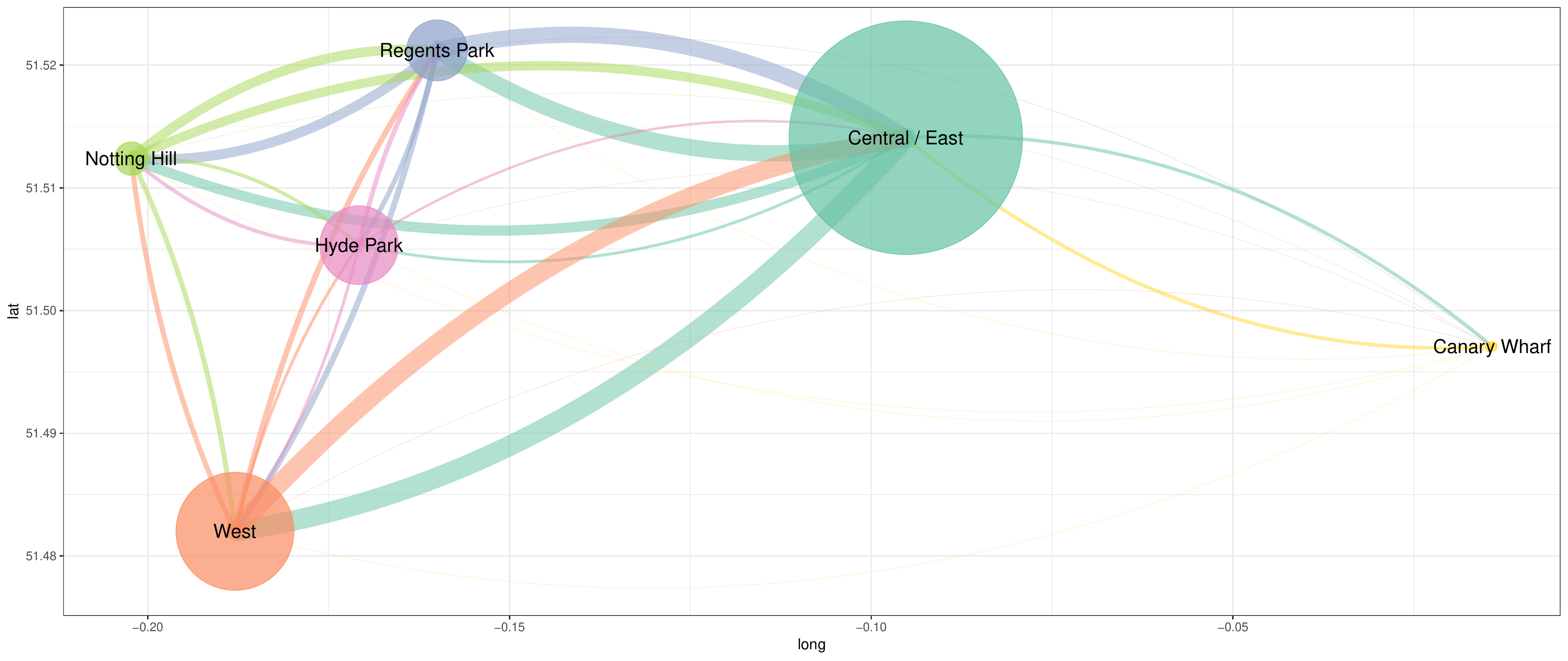}
\caption{Interactions and volume of London BSS communities: clusters are mapped at their geographic centroids. The size of point nodes and edges is scaled according to the observed flow within the community (nodes) and between the communities (edges)}
\label{fig:fig3}
\end{figure*}

\begin{algorithm}
\caption{\textit{Infomap} algorithm}\label{alg:infomap}
\begin{algorithmic}[1]
\Procedure{Infomap}{$G$}
	\State $M(N_{\alpha}) \gets M_{\alpha}$ 
	\While{$min(L(M))$}
    	\State Order $N_{1,2,...,n}$ randomly
		\For{Each $N_{\alpha}$}
			\If{$\exists M(N_{\alpha}) \gets M_{\beta}: L(M)\downarrow$}
        		\State $M(N_{\alpha}) \gets M_{\beta}$
			\Else
        		\State $M(N_{\alpha}) \gets M(N_{\alpha})$
    		\EndIf
		\EndFor
	\EndWhile
	\State \textbf{return} $M(N_{\alpha})$
\EndProcedure
\end{algorithmic}
\end{algorithm}

Dedicating a small fraction $\tau$ of the probability flow randomly links every node in the network to every other node and hence prevents the random walker from becoming stuck. 

We run the \textit{Infomap} algorithm on our dataset and return an assigned module $M_{i}$ for each station (node) $N_{\alpha}$. As this paper is the first study to apply the \textit{Infomap} algorithm in the context of station-based BSS, we compare our results to three popular modularity based methods (\textit{Greedy modularity optimization, random walks and the Louvain algorithm}), as shown in Figure 2\footnote{For our computation we use the \textit{Infomap} software package (https://github.com/mapequation/infomap, v0.19.3) and \textit{R} (v3.4.2) with the \textit{igraph} package}. The technical differences between the four approaches---discussed extensively in previous research \cite{Yang2016a}---manifest in the respective output communities. Specifically, \textit{Infomap} is the only method to detect known physical structures in London, such as Hyde Park and Canary Wharf. While the three other methods converge at four clusters, the optimal solution from \textit{Infomap} returns six modules. The first and largest community is \textit{(1)} Central and East London, accounting for more than half of the trips in the network. It borders \textit{(2)} South-West London, \textit{(3)} Regent's Park and \textit{(4)} Hyde Park clusters in the West, which are the second, third and fourth largest clusters in terms of flow. These three clusters border the fifth largest community in \textit{(5)} Notting Hill. Finally, \textit{(6)} Canary Wharf contains the least flow of any cluster and is remotely located in the South-East, only bordering the Central cluster.

\subsection{Community interactions}

\begin{figure*}[htbp]
\centering
\includegraphics[,scale=0.4]{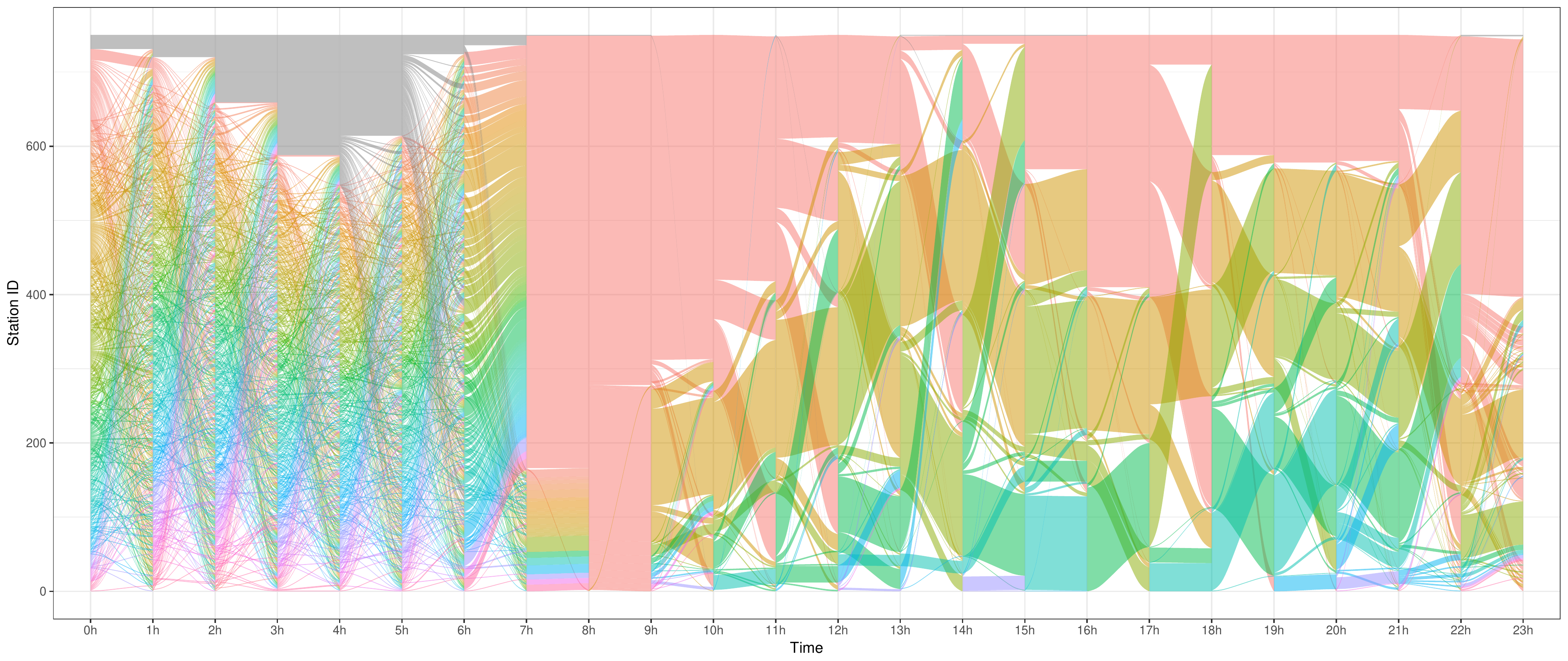}
\caption{Community evolution over time: cluster assignment for each station and hour-of-day is given using color codes (grey for no assignment). Communities are ordered and colored top-to-bottom by size.}
\label{fig:fig4}
\end{figure*}

Beyond the community detection, our results also enable insight into the interactions between the different communities. Around 75\% of the observed trips start and end within the same cluster. Nevertheless, investigating the exchange of trips between clusters provides a deeper insight into the underlying mechanics of the observed system, particularly for the smaller, more interactive communities. Our simplified community network and the flow between communities are displayed in Figure 3. The size of the communities and the links respectively highlight the observed flow. The exact numbers of trips (within, outbound and inbound) are given in Table 1. 

\begin{table}[htbp]
\caption{Summary of the community cluster characteristics and interactions}
\begin{center}
\begin{tabular}{|c|c|c|c|c|}
\hline
\textbf{}&\textbf{}&\multicolumn{3}{|c|}{\textbf{Trips}} \\
\cline{3-5} 
\textbf{Cluster} & \textbf{Stations}& \textbf{\textit{within}}& \textbf{\textit{out}}& \textbf{\textit{in}} \\
\hline
Central / East (1) & 408 & 760,404 & 112,263 & 118,146 \\
\hline
West (2) & 190 & 184,714 & 80,457 & 82,332 \\
\hline
Regents Park (3) & 71 & 48,259 & 75,618 & 71,990 \\
\hline
Hyde Park (4) & 26 & 80,354 & 63,617 & 60,289 \\
\hline
Notting Hill (5) & 35 & 17,481 & 30,443 & 29,084 \\
\hline
Canary Wharf (6) & 20 & 9,060 & 7,181 & 7,738 \\
\hline
\end{tabular}
\label{tab1}
\end{center}
\end{table}

The Central London community accounts for around 50\% of all trips in the network. Due to its large size, most trips occur within the community, while the connectivity with other clusters is particularly strong for the West London and Regents Park communities. This suggests that there might be additional hidden structures within the cluster that would allow for further simplification. The West London community behaves relatively similar, though having a higher share of trips interacting with other communities. The Regents Park and Notting Hill communities, both small in size, exhibit more interactive trips than within-cluster trips and hence suggest longer trip distance or special trip purposes. The last clusters---Hyde Park and Canary Wharf---are both small in size but nevertheless relatively isolated. For Hyde Park, a possible explanation is the specialized use of shared bicycles for leisures trips within the green-space. For Canary Wharf, this can be attributed to its remote geographic location on the Isle of Dogs, which reduces the attractiveness of bicycle trips to stations outside of the cluster. 

The community detection and interactions analysis enables us to gain novel insights into the spatial usage patterns of London's BSS. The noisy Central London cluster does not seem to exhibit explicit community structures, which hints that the usage there is less community-driven and rather might be explained using temporal analyses (e.g. commuting peak times) or destination-based approaches. The smaller, more disconnected clusters, on the other hand, suggest a strong effect of community bounds on trip-making that might be explained looking into their respective location: while the Canary Wharf cluster is located in a business area where BSS trips might serve as last-mile connections to public transport, the emergence of the Hyde Park cluster within a large public green space suggests leisure activities. This seems especially likely as our observational period is during summer time, where mild weather conditions make parks particularly attractive. On the other hand, small but interactive clusters like Notting Hill and Regents park are based around mostly residential areas and suggest the use of the BSS for commuting. As such, community detection and interaction together reflect London's physical environment and land use.

\subsection{Community dynamics}

While our previous network analyses presents novel insights into general system structures, aggregating the usage in such a large timescale results in the loss of temporal information about the way the network behaves at different times of the day. To provide a deeper understanding of the emergence and collapse of communities over time, we split our dataset into one-hour intervals for further examination. Apart from this, the methodology remains unchanged. We present the \textit{Infomap} cluster assignments for each station and all 24 hours of the day in Figure 4. The results show the evolution of the communities over the course of the day. It is apparent that during some hours---especially at nighttime---the noise in our data is considerably larger which prevents the algorithm from detecting community structures, leaving several hundreds of hardly relevant communities and even unassigned stations, due to the very low flow. This suggests a limitation of our analysis, but also relates to the general lack of observation during those unusual travel hours. From 7am, one dominant community abruptly emerges that subsumes almost all stations. This is due to the morning commuting peak, characterized by long trips connecting stations in residential areas or close to public transport facilities with the business districts. As the peak cluster disintegrates around 10am, three to four stable clusters emerge that resemble the general community structure outlined in the previous sections. This period of stability contains the vast majority of observed trips and starts and ends with the respective morning and evening peak commuting times. From around 8pm, the daytime communities slowly disintegrate towards a full collapse at midnight.

Looking at the spatial dimension, we can observe that during night-time, location does not seem to be of importance. Only starting from 10am, Central London emerges as a community, alongside clusters in Canary Wharf, Hyde Park and West London. These communities vaguely correspond to those extracted from our previous analyses (see Figure 2 and 3) and are mostly stable during daytime. Again, Canary Wharf stands out as the most isolated community, with almost all stations assigned to the same cluster from 9am to 9 pm.    

\section{Discussion and Conclusion}

Each of our three methodological sections comes with particular findings and implications relevant to BSS users, providers and public authorities. \textit{(A)} We present a new, information-theoretic method for BSS community detection that not only reflects known system features like directed links and exogenous network formation, but is also able to detect known urban built-environments like Hyde Park or Canary Wharf. Hence, we are able to infer valuable information on user behavior and the geographical boundaries of bikesharing trip-making. The findings may inform and motivate bikesharing service adoption according to the detected communities, or aim at connecting communities by incentivising users or expanding infrastructure. \textit{(B)} Our method enables us to explore bidirectional flow between communities. The imbalances in trip flows are a crucial challenge for any shared mobility scheme as they often result in vehicles getting stuck in areas characterized by high destination attractiveness and low origin attractiveness. While the communities themselves are mostly self-contained, the trips between communities and particularly their imbalances can shed new light on this issue and motivate novel relocation strategies. \textit{(C)} Lastly, we explore the emergence and collapse of communities during the course of a day, thus evaluating the noise and underlying mechanics in our system. We see that during nighttime---characterized by substantially less trips---the algorithm cannot detect clear community structures as the trips do not follow any distinct patterns. During the daytime, communities start to emerge with the clearly structured trips of the morning commuting peak and stabilize afterwards. There is some spatio-temporal fluctuations between the dominant clusters which again seem to be mostly driven by peak commuting hours. Contrarily, we find communities of remote geographic location (Canary Wharf) and specific leisure usage (Hyde Park) to be the most consistent. Beginning with the afternoon commuting peak, communities contract and eventually collapse around midnight. By exposing structure-over-time, we examine how community presence is driven by spatio-temporal dynamics. Time-sensitive events like commuting hours mostly effect certain areas---around public transport stations and business districts---while parks or leisure districts remain rather stable. These findings enable focused policies to address problems like supply shortages or congestion.

Altogether, our research reinforces the argument that BSS are inherently spatio-temporal systems of dynamic complexity. Our study also raises new questions regarding the driving factors of our observations. Further studies should extend the underlying questions from purely unsupervised learning problems to represent other layers of the urban system. For instance, research have recently shown that urban amenities \cite{Willing2017a} or weather data \cite{Reiss2015} can help with contextualizing patterns in shared mobility systems. Our findings also suggest a strong interconnection of said features with the bikesharing network, which implies that more holistic approaches are needed to draw meaningful operational and political conclusions. In future studies, this can be validated by comparing BSS to other large scale transportation networks and including further cities in those studies. Lastly, the methodological contribution of our work, while novel, could be expanded to address time-varying networks of bike stations and communities, where different motifs (loop, chain, star) and temporal evolution dynamics with extended time windows could potentially provide deeper insights into inherent relationships of spatially heterogeneous nodes (stations) or sub-networks (communities).

\section*{Acknowledgments} 

The authors gratefully acknowledge funding from the UK Engineering and Physical Sciences Research Council, the EPSRC Centre for Doctoral Training in Urban Science (EPSRC grant no. EP/L016400/1); The Alan Turing Institute (EPSRC grant no. EP/N510129/1).

%\begin{figure}[htbp]
%\centerline{\includegraphics{fig1.png}}
%\caption{Example of a figure caption.}
%\label{fig}
%\end{figure}

%\section*{References}

\bibliographystyle{IEEEtran}
\bibliography{IEEEabrv,ieee_itsc18.bib}

\end{document}